\documentclass[a4paper,11pt]{article}
\pdfoutput=1
\RequirePackage{snapshot} 
\usepackage{jheppub}
\usepackage{amsmath}
\usepackage{amssymb}
\usepackage{textcomp}
\usepackage{color}
\usepackage{graphicx}
\usepackage{hyperref}
\usepackage{xspace,slashed}
\usepackage[utf8]{inputenc}
\usepackage[caption=false]{subfig}

\usepackage{stackengine}

\usepackage{soul}

\usepackage{booktabs}

\makeatletter
\g@addto@macro\bfseries{\boldmath}
\makeatother

\newcommand*\rot{\rotatebox{90}}

\usepackage{xcolor}
\definecolor{light-gray}{gray}{0.8}

\preprint{OUTP-20-15P}

\newcommand{\CERNaff}{CERN, EP Department, CH-1211 Geneva 23, Switzerland}
\newcommand{\OXaff}{Rudolf Peierls Centre for Theoretical Physics,
  Clarendon Laboratory, Parks Road, Oxford OX1 3PU, UK
}
\title{Jet tagging in the Lund plane with graph networks}
\author[a]{Fr\'ed\'eric A. Dreyer,}%
\affiliation[a]{\OXaff}
\author[b]{Huilin Qu}%
\affiliation[b]{\CERNaff}

\abstract{%
  The identification of boosted heavy particles such as top quarks or
  vector bosons is one of the key problems arising in experimental
  studies at the Large Hadron Collider.
  In this article, we introduce LundNet, a novel jet tagging method which
  relies on graph neural networks and an efficient description of the
  radiation patterns within a jet to optimally disentangle signatures
  of boosted objects from background events.
  We apply this framework to a number of different benchmarks, showing
  significantly improved performance for top tagging compared to existing
  state-of-the-art algorithms.
  We study the robustness of the LundNet taggers to non-perturbative and
  detector effects, and show how kinematic cuts in the Lund plane can
  mitigate overfitting of the neural network to model-dependent
  contributions.
  Finally, we consider the computational complexity of this method
  and its scaling as a function of kinematic Lund plane cuts, showing
  an order of magnitude improvement in speed over previous graph-based
  taggers.
}

\begin{document}

\maketitle 
\section{Introduction}
As the Large Hadron Collider continues to explore proton collisions at
the energy frontier, an important task to search for signals of new
physics beyond the Standard Model (SM) is the identification of heavy
particles at the electroweak scale, which might emerge from decays of
yet unknown heavier particles.

An entity of particular interest in this quest is the jet, which is
essentially defined as a collimated bunch of particles with a certain
energy and direction, typically determined using a sequential
recombination algorithm~\cite{Salam:2009jx}.
One important problem is that electroweak scale particles such as
vector bosons or top quarks produced from yet heavier new states can
become sufficiently boosted such that their hadronic decays are
reconstructed as single jets.
It is therefore crucial to have efficient tools to probe the radiation
patterns within jets and determine their physical origin.
This topic has been the focus of much attention over the past decade,
with a range of approaches being developed to extract information from
a jet's
substructure~\cite{Abdesselam:2010pt,Altheimer:2012mn,Altheimer:2013yza,Adams:2015hiv,Marzani:2019hun}.
In recent years, a new generation of tools based on deep learning
models have emerged, which can achieve very high performance on
specific benchmarks~\cite{deOliveira:2015xxd,Komiske:2016rsd,Louppe:2017ipp,Kasieczka:2017nvn,Butter:2017cot,Larkoski:2017jix,Cheng:2017rdo,Macaluso:2018tck,Abdughani:2018wrw,Moreno:2019bmu,Kasieczka:2019dbj,Qu:2019gqs,Ren:2019xhp,Lim:2020igi} and provide some insights
into what kinematic variables drive the discrimination performance~\cite{Datta:2017rhs,Datta:2017lxt,Lim:2018toa,Komiske:2017aww,Komiske:2018cqr,Chakraborty:2019imr,Kasieczka:2020nyd,Agarwal:2020fpt,Chakraborty:2020yfc,Dolan:2020qkr}.
A limitation of such deep learning-based methods is the difficulty
to estimate their uncertainties, as well as their proneness to rely on
unphysical features present in the training data to achieve their high
performance, as this data is generally derived from Monte Carlo
simulations of proton collisions.

In this article, we introduce a novel method to identify jets using
graph networks.
To this end, we represent jets through their so-called Lund plane,
associating each Lund declustering with a node on the graph.
Compared with other state-of-the-art tools, our new method shows
improved performance, notably for processes with complicated
topologies such as top decays, while requiring substantially less
training time.
We will also investigate the robustness of our new tagger, and show
how through kinematic cuts on the Lund variables one can mitigate
overfitting to the model-dependent effects of Monte Carlo
simulations, reducing the reliance of the neural network on
non-perturbative contributions.
The code framework used to produce the results in this article are available as
open-source and published material
in~\cite{lundnet_code}\footnote{The code is available
  at \url{https://github.com/fdreyer/lundnet}.}.

We provide a brief review of the Lund plane for jet physics in section~\ref{sec:lund},
and describe the LundNet model in section~\ref{sec:lundnet}.
Results for a range of benchmarks are described in
section~\ref{sec:jettag}, of which a summary is given in
table~\ref{tab:benchmarks}.
The robustness and computational complexity of the models is explored
in section~\ref{sec:robustness}.
Finally, we offer our conclusions in section~\ref{sec:concl}.

\begin{table}
  \centering
  \begin{tabular}{@{}clccccc@{}}
    \toprule
     & & & AUC & & $1/\epsilon_B$ at $\epsilon_S\!=\!0.5$ & $1/\epsilon_B$ at $\epsilon_S\!=\!0.7$\\
    \cmidrule{2-7}
    & \mbox{LundNet-5}     & &  0.938  & & 609.8 & 70.4 \\[4pt]
    & \mbox{LundNet-3} & &  0.935  & & 500.0 & 61.5 \\[4pt]
    \rot{\rlap{\small{$W$ tagging}}}
    \rot{\rlap{\scriptsize{$p_t$\textgreater 500 GeV}}}
    & ParticleNet & &  0.936  & & 480.8 & 60.4 \\[2pt]
    \cmidrule{2-7}
    & \mbox{LundNet-5}     & &  0.958  & & 12500.0 & 813.2 \\[4pt]
    & \mbox{LundNet-3} & &  0.956  & & 8333.3 & 641.0 \\[4pt]
    \rot{\rlap{\small{$W$ tagging}}}
    \rot{\rlap{~\,\scriptsize{$p_t$\textgreater 2 TeV}}}
    & ParticleNet & &  0.958  & & 8333.3 & 626.4 \\[2pt]
    \cmidrule{2-7}
    & \mbox{LundNet-5} & &  0.987  & & 5000.0 & 1315.8 \\[4pt]
    & \mbox{LundNet-3} & &  0.982  & & 1785.7 & 333.3  \\[4pt]
    \rot{\rlap{\footnotesize{top tagging}}}
    \rot{\rlap{\scriptsize{$p_t$\textgreater 500 GeV}}}
    & ParticleNet & &  0.983  & & 2000.0 & 382.8  \\[2pt]
    \cmidrule{2-7}
    & \mbox{LundNet-5} & &  0.902  & & 31.6 & 12.3 \\[4pt]
    & \mbox{LundNet-3} & &  0.893  & & 27.7 & 10.7 \\[4pt]
    \rot{\rlap{\footnotesize{$q/g$ discrim.}}}
    \rot{\rlap{\scriptsize{$p_t$\textgreater 500 GeV}}}
    & ParticleNet & &  0.904  & & 34.1 & 12.9 \\[2pt]
    \bottomrule
  \end{tabular}
  \caption{Benchmarks of several jet tagging algorithms for a range of
    processes. The first column gives the area under the ROC curve,
    and the later two show the background rejection at two different
    signal efficiencies, 50\% and 70\% respectively. In each case,
    larger values indicate better performance.}
  \label{tab:benchmarks}
\end{table}

\section{Jets in the Lund plane}
\label{sec:lund}

The models we introduce in this article rely on the Lund
plane~\cite{Andersson:1988gp}.
This representation provides a useful mapping of the emission
phase-space to a two dimensional plane representing the angle and
transverse momentum of a given emission with respect to its emitter,
and which is often used in discussions of resummations of large
logarithms in perturbation theory or of Monte Carlo parton showers.
Each emission then creates an additional triangular leaf corresponding to
the phase space for further emissions.
%
%
%
It was shown in recent work that the Lund plane provides a useful
basis to achieve an efficient description of the clustering sequence
of a jet, containing a rich set of information about its substructure,
with notable potential for jet tagging~\cite{Dreyer:2018nbf}.
The Lund jet plane allows for a visual representation of the clustering history
of a jet.
This systematic encoding of a jet's radiation patterns can be measured
experimentally~\cite{Aad:2020zcn}, allowing for comparisons between
theoretical predictions and experimental data~\cite{Lifson:2020gua}
and with potential for constraining general purpose Monte Carlo event
generators~\cite{Dasgupta:2020fwr}.
\begin{figure}
  \centering
  \hspace{-18mm}\includegraphics[width=\textwidth]{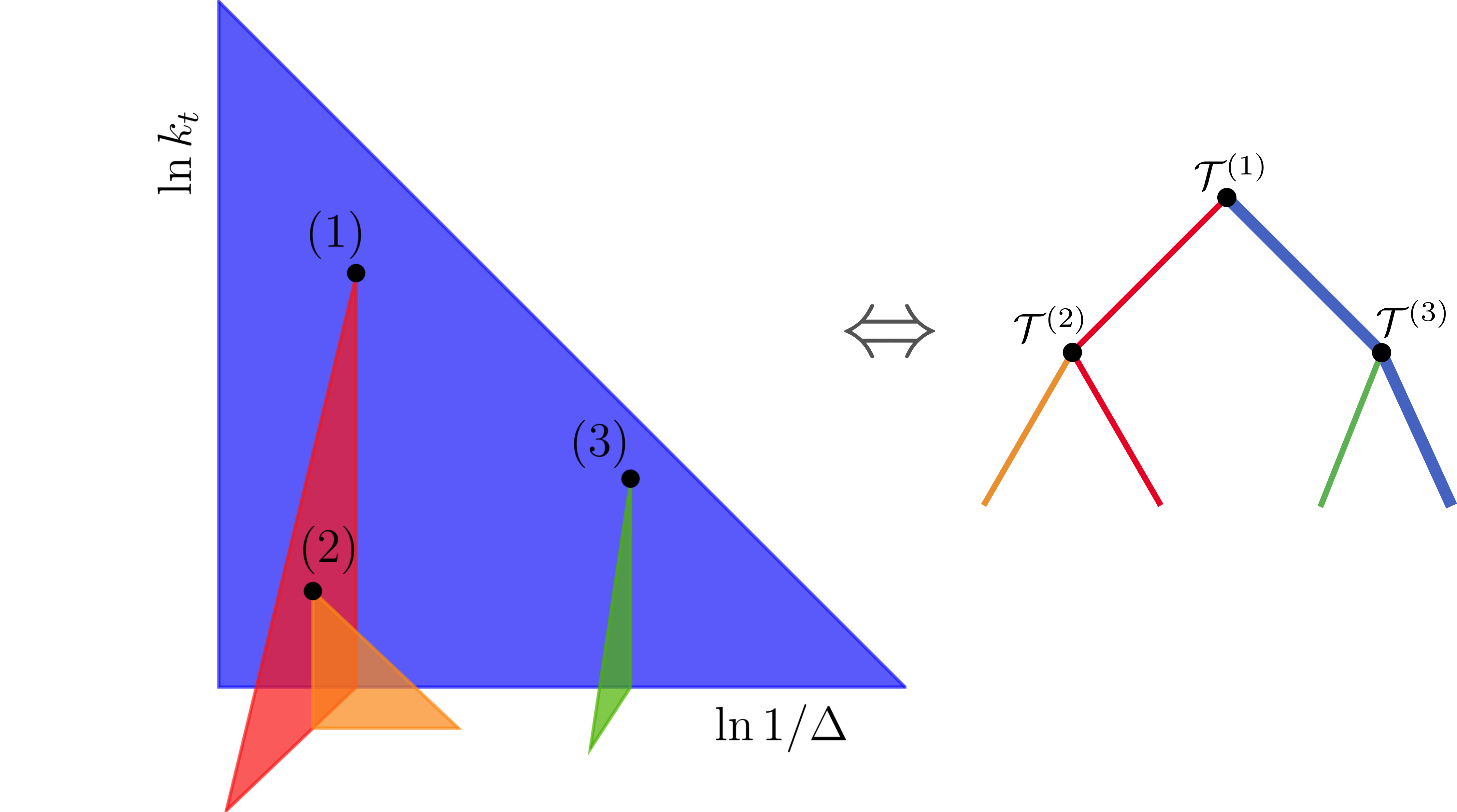}%
  \caption{The Lund plane representation of a jet (left) where each
    emission is positioned according to its $\Delta$ and $k_t$
    coordinates, and the corresponding mapping to a binary Lund tree of
    tuples (right). The thick blue line represents the primary
    sequence of tuples $\mathcal{L}_\text{primary}$.}
  \label{fig:lundplane}
\end{figure}

The Lund plane is obtained by first reclustering a jet's constituents
with the Cambridge/Aachen (CA)
algorithm~\cite{Dokshitzer:1997in,Wobisch:1998wt}, which sequentially
identifies and combines the pair of particles $a$ and $b$ closest in
rapidity $y$, a measure of relativistic velocity along the beam axis,
and azimuthal angle $\phi$ around the same axis, i.e.\ minimising
$\Delta^2 = (y_a - y_b)^2 + (\phi_a - \phi_b)^2$.
We then iterate over this clustering sequence, starting from the full
jet and proceeding by:

\begin{enumerate}
\item Declustering the current (pseudo)jet into two transverse momentum
  ordered pseudojets $a$ and $b$ such that $p_{t,a} > p_{t,b}$, and
  where we consider $b$ to be the emission of the $(a+b)$ emitter.
\item Determining a number of kinematic variables associated with the
  declustering step $i$, which we denote as a tuple $\mathcal{T}^{(i)}$
  \begin{equation}
    \label{eq:tuple}
    \mathcal{T}^{(i)} = \{k_t, \Delta, z, m, \psi\}\,.
  \end{equation}
  Here $k_t= p_{t,b}\Delta$ is the transverse momentum of emission $b$
  with respect to its emitter in the limit where $p_{t,b}\ll p_{t,a}$,
  $\Delta$ is the previously defined rapidity-azimuth distance, 
  $z=p_{t,b}/(p_{t,a}+p_{t,b})$
  is the momentum fraction of the softer subjet $b$, $m$ is
  the invariant mass of the $(a+b)$ pair, and
  $\psi=\tan^{-1}\big(\tfrac{y_b-y_a}{\phi_b-\phi_a}\big)$ is the
  azimuthal angle around subjet $a$'s axis.
\item Repeating this procedure for pseudojets $a$ and $b$ if they
  contain more than one particle.
\end{enumerate}
This procedure produces a binary Lund tree with a tuple of variables
$\mathcal{T}^{(i)}$ for each node $i$ of the Lund tree, as shown in
figure~\ref{fig:lundplane}.
The first two elements of the tuple provide the coordinates in the Lund
plane of the corresponding splitting, and the remaining ones provide
complementary kinematic information.
A subset of this tree of particular significance is the primary list
of tuples $\mathcal{L}_\text{primary}$ containing the kinematic
variables of each splitting along the primary branch of the tree,
i.e.\ following only the pseudojet with larger transverse momentum in
step 3.\ of the algorithm above, corresponding to points on the
blue primary plane in figure~\ref{fig:lundplane}.
The primary Lund sequence can be used notably for two-dimensional visual
representations of the radiation patterns in a
jet~\cite{Dreyer:2018nbf,Aad:2020zcn,Andrews:2018jcm}.

\begin{figure}
  \centering
  \includegraphics[width=0.5\textwidth]{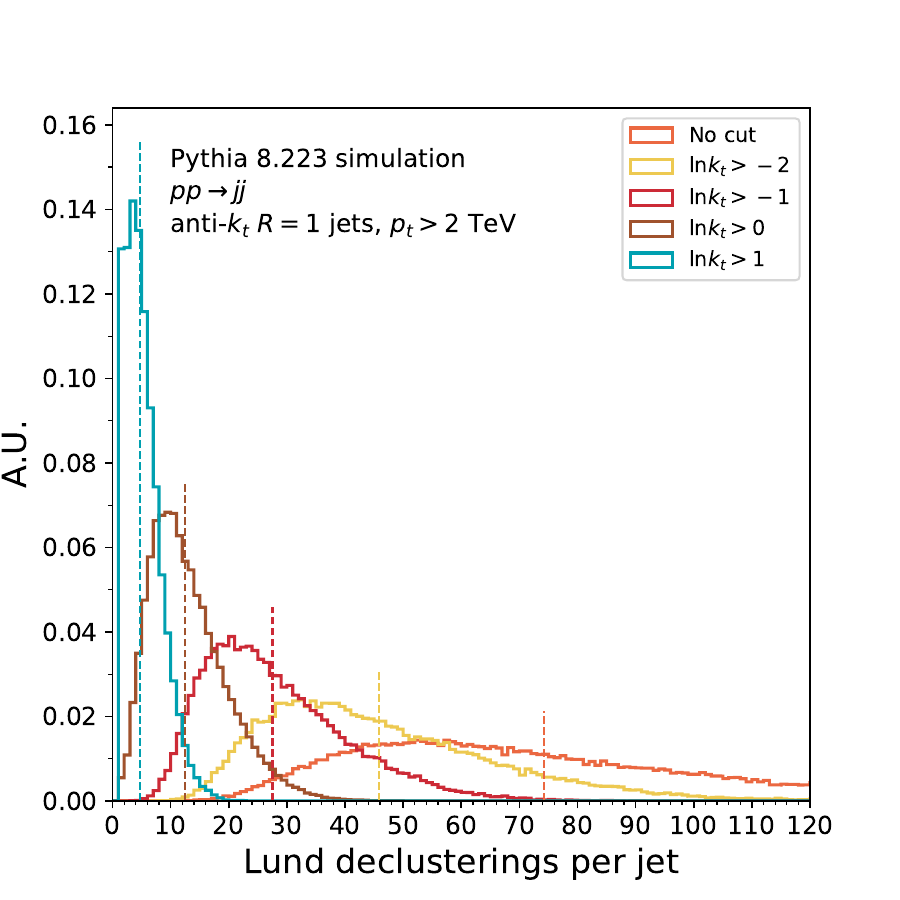}%
  \includegraphics[width=0.5\textwidth]{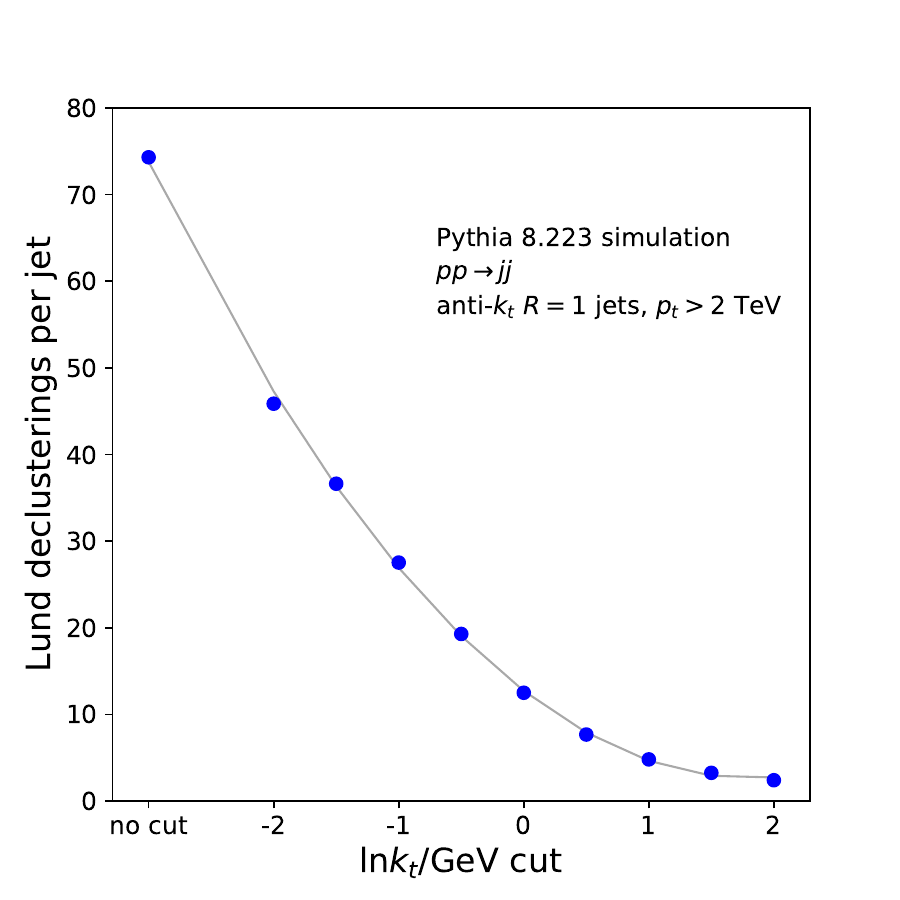}%
  \caption{Left: Distribution of the number of Lund declusterings per jet, for
    different choices of $k_t$ cuts in the Lund plane.
    The dashed lines indicate the mean of each distribution.
    Right: Average of the number of nodes per jet as a function of the
    $k_t$ cut.}
  \label{fig:nodes}
\end{figure}

Corrections to the Lund plane originating from non-perturbative
hadronisation effects affect the low $k_t$ region of the plane.
One can therefore limit the dependence on the non-perturbative region
of any model trained on Lund declusterings by removing emissions that
fall below a certain transverse momentum $k_t$ threshold.
In figure~\ref{fig:nodes} (left), we show the distribution of the number of
Lund declusterings per jet for several choices of $k_t$ cut in 2 TeV
QCD jets simulated using \texttt{Pythia} 8.223~\cite{Sjostrand:2014zea}.
The mean of each distribution is indicated as a dashed line.
An additional benefit of a $k_t$ threshold is that even for small cut
values the number of nodes per jet is significantly reduced, and
therefore correspondingly so the computational cost of training a
machine learning model on these inputs.
The right-hand side of figure~\ref{fig:nodes} shows the average number
of nodes per jet as a function of the $k_t$ cut, which decreases
quadratically as the cut is increased.

\section{LundNet Models}
\label{sec:lundnet}

\begin{figure}
  \centering
  \includegraphics[width=0.95\textwidth]{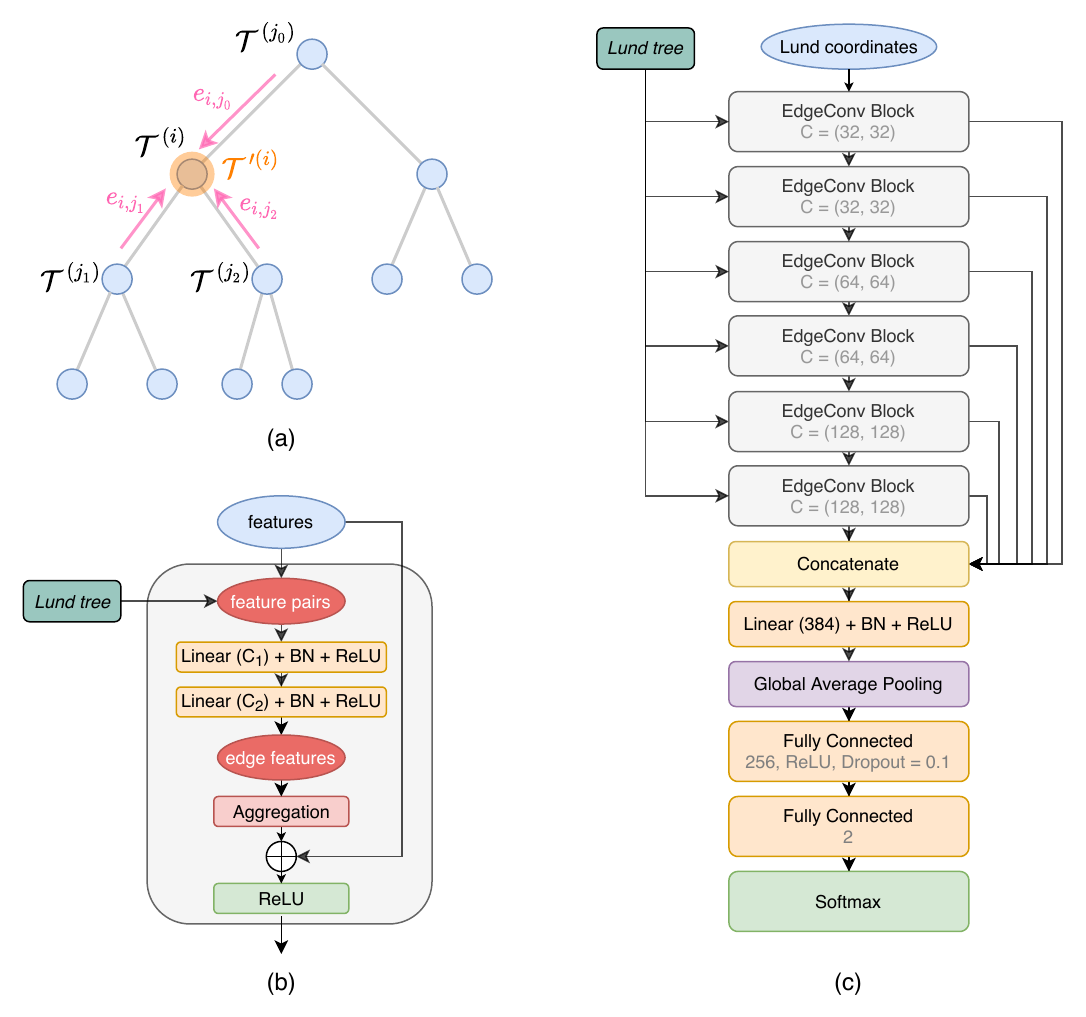}%
  \caption{(a) Illustration of the EdgeConv operation on a node of the Lund tree. 
  (b) Architecture of the EdgeConv block used in the LundNet model. 
  (c) Architecture of the LundNet model.}
  \label{fig:network}
\end{figure}

The Lund plane encodes a rich set of information of the substructure
and radiation patterns of a jet, therefore serving as a natural input
to machine learning models for jet physics. The use of Lund planes for
jet tagging was first proposed in Ref.~\cite{Dreyer:2018nbf} where
log-likelihood and deep learning models are applied, and good
performance was observed for tagging boosted electroweak
bosons.
However, the main focus of Ref.~\cite{Dreyer:2018nbf} was the primary Lund plane, which
inevitably leads to some loss of information due to the omission of
the secondary and tertiary splittings.
In this article, we propose LundNet, a new deep learning model capable
of digesting the full Lund plane. Graph neural networks are used in
this model to better exploit the structural information associated
with the Lund plane representation of a jet, leading to significantly
improved performance on a range of jet tagging benchmarks.

The LundNet model starts with transforming the Lund tree into a graph, where 
each node corresponds to a Lund declustering and 
carries the tuple of kinematic variables $\mathcal{T}^{(i)}$ as its input features, 
and bidirectional edges are formed following the structure of the 
Lund declustering tree.
The graph network architecture is adapted from the
ParticleNet~\cite{Qu:2019gqs} model, with the EdgeConv operation
proposed in Ref.~\cite{DGCNN} as a core step.
Figure~\ref{fig:network}(a) illustrates 
how EdgeConv operates for one node (the highlighted one) in the Lund tree. 
It consists of two steps: First, a shared multi-layer perceptron (MLP) is applied 
to each of its incoming edges, using features of the node pair connected by 
the edge as inputs, and produces a learned ``edge feature''.
As the Lund tree is a binary tree, there are only up to three edges
for each node, which do not require a nearest-neighbour search,
therefore the computational cost is much lower than for the
ParticleNet model.
As shown in figure~\ref{fig:network}(b), we use two layers for this
shared MLP, each consisting of a linear layer followed by a batch
normalization (BN)~\cite{DBLP:journals/corr/IoffeS15} and a ReLU
activation~\cite{glorot2011deep}.
Then, an aggregation step is performed for the node by taking an
element-wise average of the learned edge features of all the incoming
edges.
A shortcut connection~\cite{he2016deep} is also added to take the
original node features into account directly, and the node feature is
then updated to the new value.
This operation is performed for all the nodes using the same shared
MLPs, therefore updating all the node features but keeping the graph
structure unchanged.

The architecture of the LundNet model is shown in figure~\ref{fig:network}(c). 
We stack six such EdgeConv blocks to form a deep graph network. 
The number of channels of the MLPs are 
$(32, 32)$, $(32, 32)$, $(64, 64)$, $(64, 64)$, $(128, 128)$ and $(128, 128)$
for the six EdgeConv blocks, respectively. 
Outputs from these EdgeConv blocks are concatenated per node and further processed by 
another MLP with 384 channels to better aggregate features learned at different stages. 
A global average pooling is applied afterwards to read out information from all nodes 
in the graph. This is followed by a fully connected layer with 256 units 
and a dropout layer with a drop probability of 0.1, before the final 
classification output. 

The LundNet model uses the Lund kinematic variables defined in equation~(\ref{eq:tuple}) as 
the input node features. Two variants of the LundNet models are investigated in this 
article. The first one uses all five Lund variables, 
\begin{equation}
  \label{eq:lundnet5-input}
(\ln k_t, \ln\Delta, \ln z, \ln m, \psi)\,
\end{equation}
as input features to extract as much information as possible from the Lund plane to 
maximize the jet tagging performance and is referred to as \mbox{LundNet-5}. The second one 
uses only three Lund variables, 
\begin{equation}
  \label{eq:lundnet3-input}
  (\ln k_t, \ln\Delta, \ln z)\,
\end{equation}
and is referred to as \mbox{LundNet-3}. The removal of the $\ln m$ and $\psi$ variables 
significantly increases the resilience of the model to non-perturbative effects 
at only a small cost of the performance, as will be discussed 
in Section~\ref{sec:robustness}.

We implement the LundNet model with the Deep Graph Library~0.4.3 \cite{wang2019dgl} using 
the PyTorch~1.7 \cite{NEURIPS2019_9015} backend. The training is performed on a 
Nvidia GTX 1080 Ti graphics card with a minibatch size of 256. The Adam 
optimizer \cite{DBLP:journals/corr/KingmaB14} is used to minimize the cross entropy 
loss. The training is 
performed for 30 epochs, with an initial learning rate of 0.001, and 
subsequently lowered by a factor of 10 after the 10th and the 20th epochs. 
A snapshot of the model is saved at the end of each epoch, and the
model snapshot showing the best accuracy on the validation dataset is 
selected for the final evaluation.

\section{Jet tagging in the Lund plane}
\label{sec:jettag}

Let us now turn to a detailed evaluation of our models for the
identification of several hallmark signals at the LHC.
We will look at four different benchmarks: the tagging of boosted
electroweak $W$ for two different transverse momentum cuts, the
tagging of top quarks, and the discrimination between quark and gluon
jets.
The data samples consist in each benchmark of 1.2m signal and
background jets simulated through the corresponding process with
\texttt{Pythia} 8.223~\cite{Sjostrand:2014zea}, with an equal split
between signal and background events.
Events are generated at hadron-level with underlying event turned on,
but without including detector effects or the presence of additional
pile-up collisions.
A subset of 100k jets each are used as validation and test data, with
the same number of signal and background events in both samples.
All data sets are taken from Ref.~\cite{Carrazza:2019efs,groomRL_data}.
Jets are clustered using the anti-$k_t$
algorithm~\cite{Cacciari:2008gp,Cacciari:2011ma} with a radius $R=1.0$
using \texttt{FastJet}~3.3.2, and are required to pass a selection
cut, with transverse momentum $p_t > 500$ GeV or $p_t > 2$ TeV as
indicated, and rapidity $|y|<2.5$.
In each event, only the two jets with the highest transverse momentum
are considered, and are saved as training data if they pass the
selection cuts.
A summary of these benchmarks is given in table~\ref{tab:benchmarks}.

\subsection{$W$ tagging}
\label{sec:wtag}

\begin{figure}
  \centering
  \includegraphics[width=0.75\textwidth,page=1]{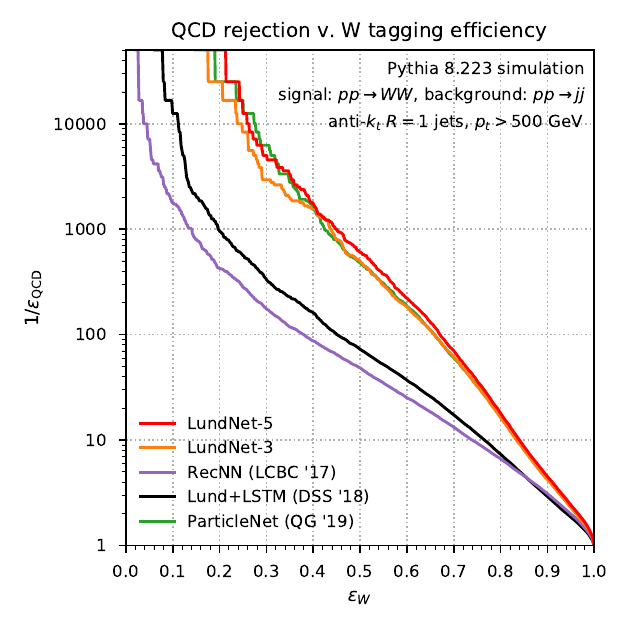}
  \caption{Background rejection $1/\epsilon_\text{QCD}$ versus signal
    efficiency $\epsilon_W$ for $W$ jet tagging with transverse
    momentum $p_t > 500$ GeV.}
  \label{fig:tagging-W}
\end{figure}

We start by considering the identification of hadronically decaying
$W$ bosons, one of the key objects commonly appearing in high energy
proton collisions.
The signal data is obtained from 600k jets passing the selection
cuts and simulated using the $pp\rightarrow WW$ process, where the $W$
bosons are decayed hadronically.
The background consists of the same number of QCD jets simulated
through a sample of $pp\rightarrow jj$ events.
Training of the neural network weights for every model is performed
using 500k of the $W$ and background samples each.
At the end of each epoch, the performance is monitored on a separate
validation sample consisting of 100k jets.
The final performance of each model is then evaluated using a further
independent sample of 100k jets with an equal number of signal and
background events.

In fig.~\ref{fig:tagging-W} we show for each model the background
rejection $1/\epsilon_\mathrm{QCD}$ against the signal efficiency
$\epsilon_W$ for $W$ bosons, for jets passing a transverse momentum cut of $p_t > 500$ GeV.
Better performance translates to curves that achieve a higher background
rejection for a given signal efficiency, i.e.\ which are closer to the
top right corner of the figure.
We compare the \mbox{LundNet-3} and \mbox{LundNet-5} models with three recent
benchmarks: the ParticleNet model introduced in~\cite{Qu:2019gqs}, the RecNN
model from~\cite{Louppe:2017ipp} and the Lund+LSTM model from the original Lund
plane paper~\cite{Dreyer:2018nbf}, which uses an LSTM network on the primary Lund sequence.
Both the RecNN and the Lund+LSTM models, while superior to heuristic
substructure algorithms, are vastly outperformed by all of the graph
based methods considered.
The \mbox{LundNet-3} model is able to achieve about the same signal
purity as ParticleNet, but can be trained in substantially less time,
as will be discussed in more detail in section~\ref{sec:complexity},
and takes only a small 3-dimensional input for each declustering node
in the Lund plane.
By including more kinematic information, the \mbox{LundNet-5} model
is able to provide a slightly higher performance, but as we will see
in section~\ref{sec:robustness}, this comes at the price of being less
robust to non-perturbative effects than its lower-dimensional
counterpart.

\begin{figure}
  \centering
  \includegraphics[width=0.75\textwidth,page=2]{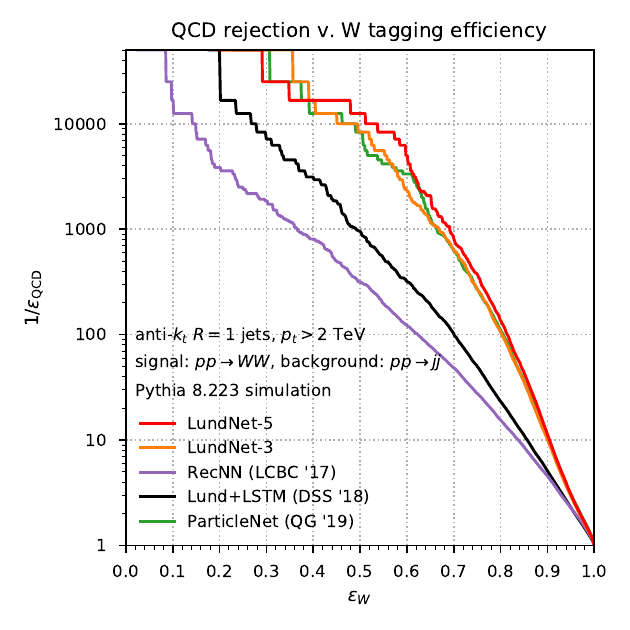}
  \caption{Background rejection $1/\epsilon_\text{QCD}$ versus signal
    efficiency $\epsilon_W$ for $W$ jet tagging with transverse
    momentum $p_t > 2$ TeV.}
  \label{fig:tagging-W2}
\end{figure}

In figure~\ref{fig:tagging-W2}, we show the same process but with a
transverse momentum selection cut of $p_t > 2$ TeV for the jets.
Here we can observe roughly the same qualitative behaviour as at lower
transverse momentum, but with the \mbox{LundNet-5} model now clearly
outperforming the remaining taggers even at high signal efficiencies.
At higher transverse momentum, the peak in the Lund plane associated
with the $W$ splitting, and the corresponding depletion associated
with the colour-singlet nature of the $W$, become more
distinguishable.
The Lund+LSTM model, which relies purely on the primary Lund sequence,
also shows a strong performance, although it is still lags
significantly behind all the graph-based approaches.

\subsection{Top tagging}
\label{sec:top}

\begin{figure}
  \centering
  \includegraphics[width=0.75\textwidth,page=3]{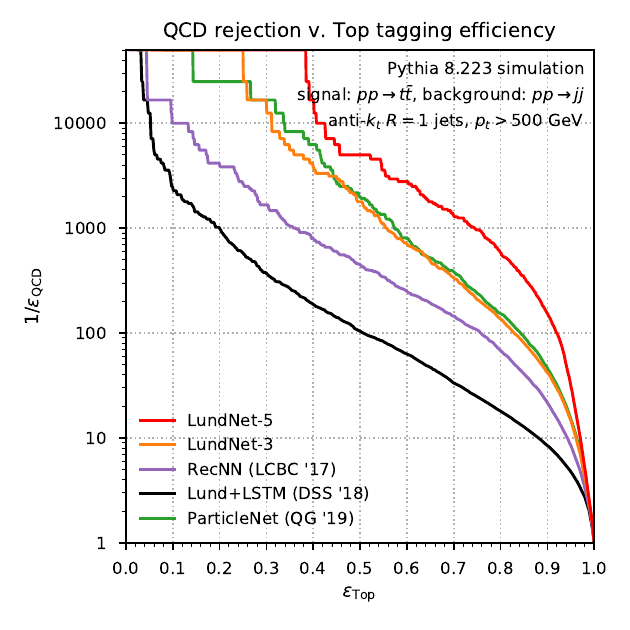}
  \caption{Background rejection $1/\epsilon_\text{QCD}$ versus signal
    efficiency $\epsilon_\text{Top}$ for top jet tagging with transverse
    momentum $p_t > 500$ GeV.}
  \label{fig:tagging-top}
\end{figure}

We now turn to the identification of jets originating from top quark
decays.
Top quarks are of particular interest at the LHC, interacting strongly
with the Higgs boson and providing a valuable avenue in searches for
new physics, as well as being the only quarks to decay before
hadronising.
Here the signal data is obtained from the $pp\rightarrow t\bar{t}$
process in \texttt{Pythia}~8.223, where the top quarks decay to hadrons and
the jets are required to pass a 500 GeV transverse momentum cut.
The background QCD jets are identical to the ones used in
figure~\ref{fig:tagging-W}.
Each model is again trained using 500k signal and 500k background
jets, with further validation and testing samples that are both one tenth
the size of the training data.

In fig.~\ref{fig:tagging-top}, we show the QCD background rejection as
a function of the top efficiency.
In this case, the Lund+LSTM model does not perform as well as RecNN.
This is to be expected, as it was designed for one or two-pronged jet
identification and uses only information from the primary Lund
declustering sequence.
It therefore contains information about the structure of only one of
the initial decay products of the original top quark, limiting the
performance that can be achieved without input from secondary planes.
It is however interesting to see that in this process with more
complex topology, the \mbox{LundNet-5} model provides a substantial
performance gain over existing state-of-the-art methods such as
ParticleNet.
This is due to the nature of its input, which contains already
high-level kinematic information about the radiation patterns of the
jet, making it much simpler for the neural network to learn how to
distinguish signals with more involved signatures.
Thus the \mbox{LundNet-3} model achieves almost the same signal
purity as the ParticleNet algorithm, despite having as input only a
reduced 3-tuple of kinematic variables per node and taking about an
order of magnitude less time to train.
Interestingly, the performance gap between the two LundNet taggers is
entirely due to the addition of the subjet mass and azimuthal angle
$\psi$ to the input features of each declustering for the
\mbox{LundNet-5} model.

\subsection{Quark/Gluon discrimination}
\label{sec:qg}

\begin{figure}
  \centering
  \includegraphics[width=0.75\textwidth,page=4]{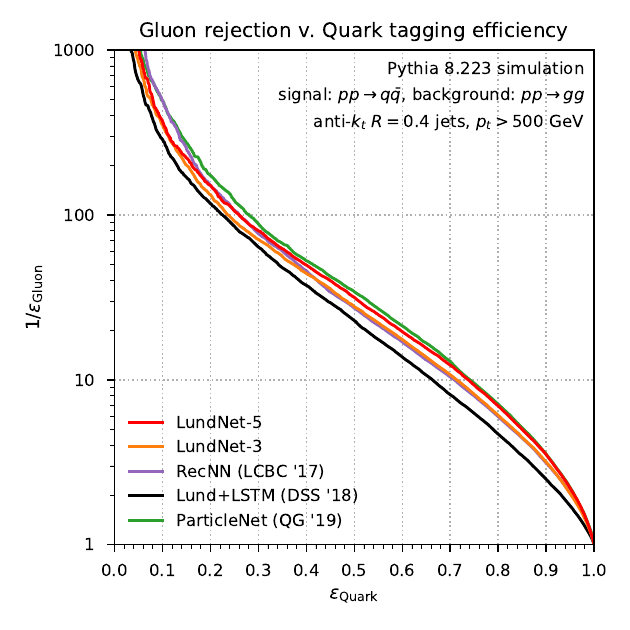}
  \caption{Background rejection $1/\epsilon_\text{Gluon}$ versus
    signal efficiency $\epsilon_\text{Quark}$ for quark/gluon
    discrimination between $R=0.4$ anti-$k_t$ jets with transverse momentum
    $p_t > 500$ GeV.}
  \label{fig:tagging-qg}
\end{figure}

Our final benchmark considers the discrimination between quark and
gluon initiated jets, a core challenge in collider physics which has
seen much research in recent
years~\cite{Gallicchio:2011xq,Larkoski:2013eya,Larkoski:2014pca,Bhattacherjee:2015psa,Komiske:2016rsd,Frye:2017yrw,Metodiev:2018ftz,Larkoski:2019nwj}.
For this study, we consider a signal sample of 500k quark-initiated
jets obtained through the $q\bar{q}\rightarrow q\bar{q}$ process in
\texttt{Pythia}~8.223, while the background is obtained from
$gg\rightarrow gg$ events.
The jets are clustered with an anti-$k_t$ algorithm with radius
$R=0.4$ and are again required to pass a transverse momentum $p_t > 500$ GeV
and rapidity $|y|<2.5$ selection cut.

The gluon-jet rejection as a function of the quark-jet efficiency is
shown in figure~\ref{fig:tagging-qg}.
In this case there is not as large a hierarchy between models, with
the Lund+LSTM model performing somewhat below the competing
approaches.
ParticleNet has a slight edge over the other algorithms at small quark
efficiencies, but is indistinguishable from the \mbox{LundNet-5} tagger at high
efficiency.
The \mbox{LundNet-3} and RecNN models show similar performance at
high efficiency, with RecNN providing slightly higher gluon rejection
at lower quark efficiencies.

\section{Robustness study}
\label{sec:robustness}

We will now investigate the robustness of the different models we
considered in our benchmarks.
To this end we will consider three axes: their resilience to
non-perturbative effects, their resilience to detector effects, and
the complexity and computational cost of each tagger.

\subsection{Non-perturbative effects}
\label{sec:np-effects}

\begin{figure}
  \centering
  \includegraphics[width=0.85\textwidth,page=3]{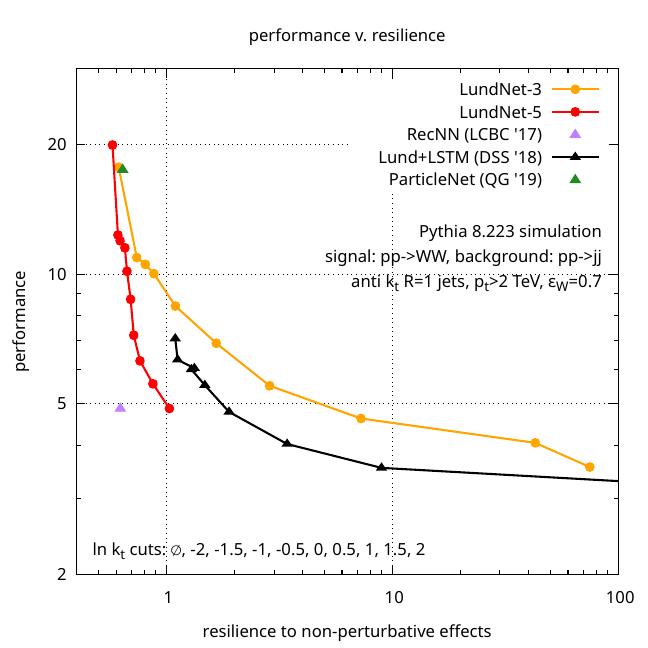}
  \caption{Performance $\tfrac{\epsilon_W}{\sqrt{\epsilon_\text{QCD}}}$ versus resilience to non-perturbative effects.}
  \label{fig:perf-resi-np}
\end{figure}

Beyond its raw performance, it is important for practical applications
that a tagger be relatively robust to model-dependent non-perturbative
effects.
To carry out studies of sensitivity to non-perturbative effects, we
compare performance between a data sample of both 50k signal and
background jets produced at parton level, and a sample obtained with
hadronisation and underlying event models turned on in the event
generator.
The same model, trained on hadron-level data, is evaluated on 
both samples for the comparison.
For this study, we use the same 2 TeV $W$ jet sample as was used in
section~\ref{sec:wtag} as well as the corresponding models shown in
figure~\ref{fig:tagging-W2}, which are now used to label jets from 
both parton and hadron-level data.

\begin{figure}
  \centering
  \includegraphics[width=0.5\textwidth,page=1]{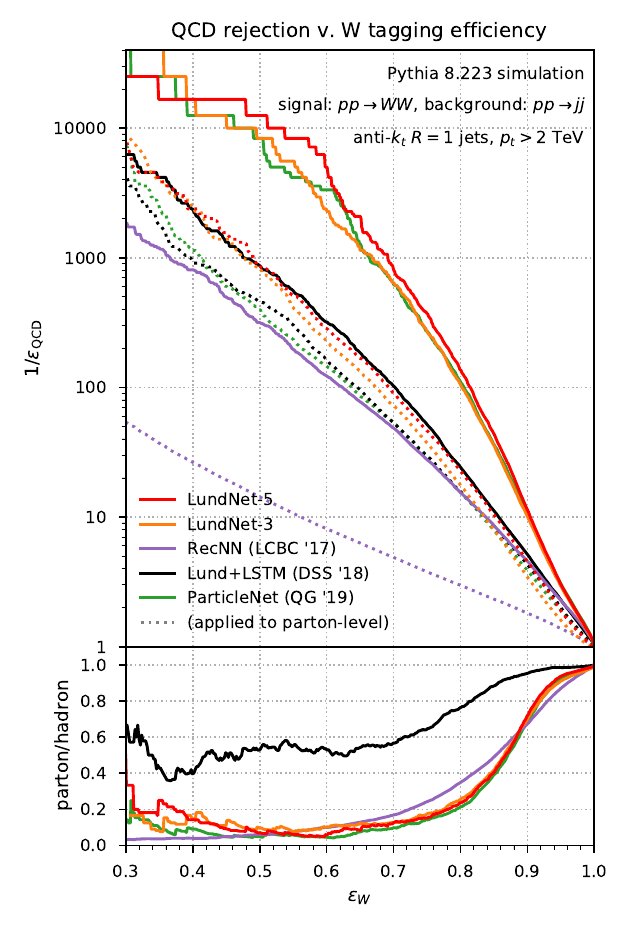}%
  \includegraphics[width=0.5\textwidth,page=2]{figures/plot-rocs-np.pdf}%
  \caption{Background rejection as a function of $W$ tagging efficiency. Dotted lines indicate a $W$ tagger applied on parton-level data.}
  \label{fig:roc-np}
\end{figure}

%
Figure~\ref{fig:perf-resi-np} shows the robustness of the tagger 
in conjunction with its performance.
This robustness is measured through the resilience $\zeta_\text{NP}$~\cite{Bendavid:2018nar},
calculated using both the efficiency on the hadron-level sample, $\epsilon$, and that
on the parton-level sample, $\epsilon'$
\begin{equation}
  \label{eq:resi}
  \zeta_\text{NP} = \left(\frac{\Delta \epsilon_W^2}{\langle\epsilon\rangle^2_W}
    + \frac{\Delta \epsilon_\mathrm{QCD}^2}{\langle\epsilon\rangle^2_\mathrm{QCD}}
  \right)^{-1/2}\,,
\end{equation}
where $\Delta \epsilon = \epsilon - \epsilon'$ and
$\langle\epsilon\rangle=1/2 \left(\epsilon + \epsilon' \right)$.
The efficiencies are obtained with a fixed cut corresponding to a signal 
efficiency $\epsilon_W=70\%$ on the hadron-level sample.
The curves in figure~\ref{fig:perf-resi-np} are obtained by increasing
a transverse momentum cut on the $k_t$ variable of the Lund plane,
progressively removing declustering nodes that fall below the cut.
Each curve starts on the upper left of figure~\ref{fig:perf-resi-np},
with a model trained without any cuts on the Lund plane, and ends in
the lower right part of the figure with a model trained with a
transverse momentum cut $\ln k_t/\text{GeV} > 2$ that has higher
resilience but lower performance due to the removal of parts of the
Lund tree.
We can observe that despite their good performance, the ParticleNet
and RecNN models have very little resilience to non-perturbative
effects, and have no handles through which such robustness can be
consistently imposed.
Somewhat surprisingly, the \mbox{LundNet-5} also offers relatively poor robustness
to non-perturbative effects.
This is due to its higher dimensional input state, which allows the
neural network to extrapolate some information on emissions in the
non-perturbative regime despite the presence of a transverse momentum
cut.
In particular, the mass variable present in the input of the
\mbox{LundNet-5} model contains information about soft wide-angle
emissions further down the clustering tree, as these will increase the
mass of the subjet even if they are then removed after failing a
$k_t$ cut.
In contrast, the \mbox{LundNet-3} model becomes very resilient to
non-perturbative effects as the transverse momentum cut is
increased, outperforming the Lund+LSTM model by a factor two for the
same resilience value.

In figure~\ref{fig:roc-np} (left), we show the ROC curve for each model
trained on the hadron-level $W$ data, with the ROC curve obtained on
the parton-level data shown as a dotted line.
The lower panel provides the ratio between the parton-level ROC curve
and the hadron-level one.
The right-hand side of figure~\ref{fig:roc-np} gives the ROC curve of
the \mbox{LundNet-3} models obtained for several choices of the
$\ln k_t$ transverse momentum cut applied on the Lund tree.
Here we can observe the improved resilience as the $k_t$ cut is
increased, with the $\ln k_t/\text{GeV} > 1$ model providing almost
the same performance at parton and hadron level, albeit at the cost of
a factor 20 in background rejection when compared to the unconstrained
model.

\subsection{Detector effects}
\label{sec:detector}

\begin{figure}
  \centering
  \includegraphics[width=0.85\textwidth,page=3]{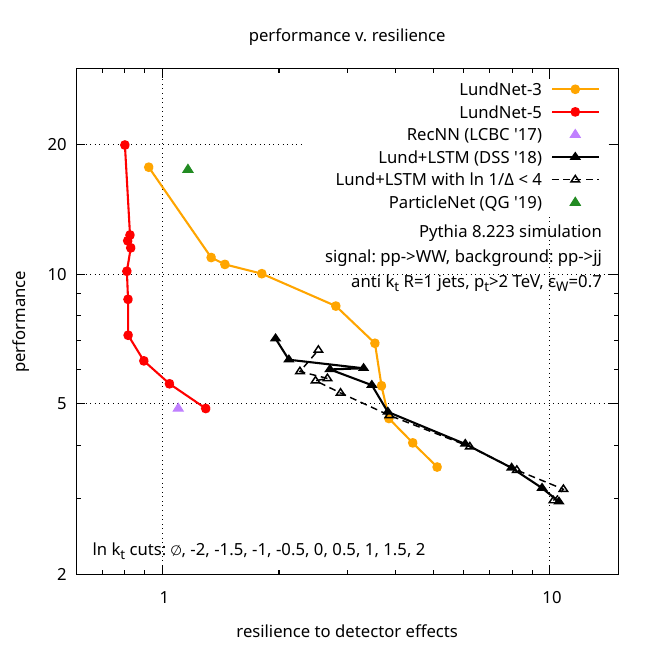}
  \caption{Performance $\tfrac{\epsilon_W}{\sqrt{\epsilon_\text{QCD}}}$ versus resilience to detector effects.}
  \label{fig:perf-resi-detect}
\end{figure}

Let us now turn to the impact of detector effects on the model robustness.
To this end, we create a sample of 100k 2 TeV $W$ and QCD jets using
\texttt{Pythia}~8.223, including fast detector simulation with
\texttt{Delphes}~v3.4.1 using the
\texttt{delphes\_card\_CMS\_NoFastJet.tcl} card to simulate both
detector effects and particle flow
reconstruction~\cite{deFavereau:2013fsa}.
The effects of detector granularity are then partially mitigated by
applying a subjet-particle rescaling
algorithm~\cite{Dreyer:2018nbf,Son:2012mb}, where the Delphes
particle-flow objects in a jet are reclustered into subjets using a CA
algorithm with $R_h=0.12$ and rescaling the particle flow
charged-particle and photons by a factor
\begin{equation}
  \label{eq:SPRA1}
  f = \frac{\sum_{i\in\text{subjet}} p_{t,i}}{\sum_{i\in\text{subjet}(h^{\pm},\gamma)} p_{t,i}}.
\end{equation}
before discarding neutral hadron candidates. The resulting particles
of all subjets are then reclustered into a single jet on which the
Lund tree can be measured.

\begin{figure}
  \centering
  \includegraphics[width=0.5\textwidth,page=3]{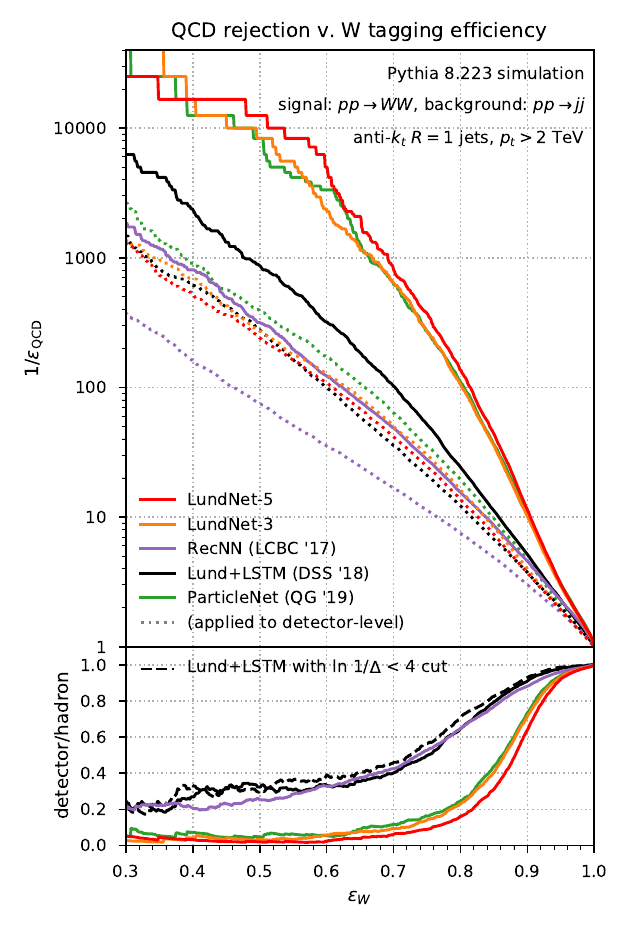}%
  \includegraphics[width=0.5\textwidth,page=2]{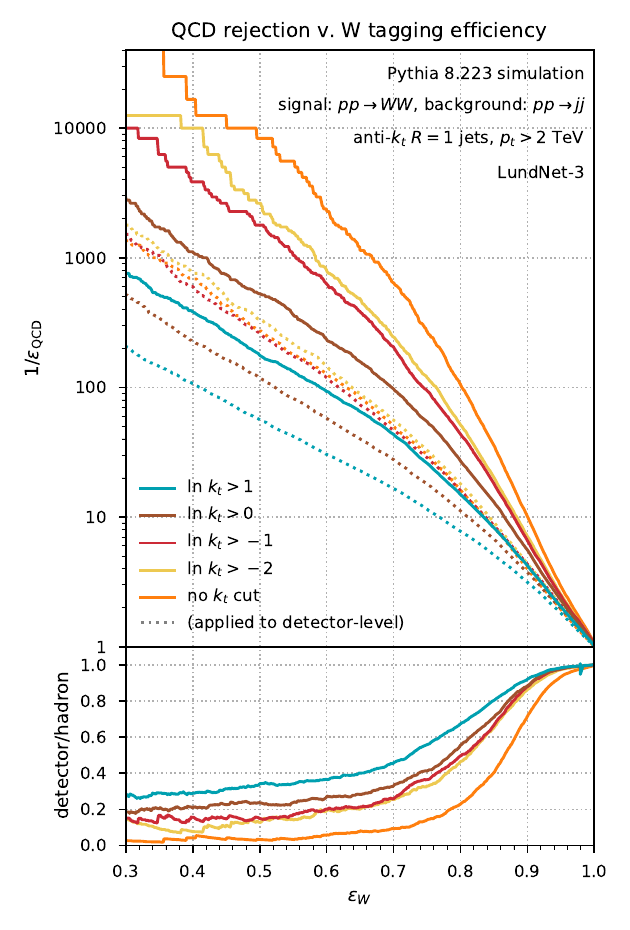}%
  \caption{Background rejection as a function of $W$ tagging efficiency. Dotted lines indicate a $W$ tagger applied on detector-level data.}
  \label{fig:roc-detec}
\end{figure}

Applying the 2 TeV $W$ taggers trained in section~\ref{sec:wtag} on
this sample, we can now compute an index of resilience to detector
effects $\zeta_\text{D}$ in the same way as was done for
non-perturbative effects, but taking now $\epsilon'$ in
equation~(\ref{eq:resi}) to be the detector-level efficiency.
We show in figure~\ref{fig:perf-resi-detect} the resilience as a
function of performance $\epsilon_W / \sqrt{\epsilon_\text{QCD}}$ for a
signal efficiency of $\epsilon_W=70\%$ on the hadron-level data.
One can observe here that while for high performance, good resilience
can be achieved, a transverse momentum cut in the Lund plane does not
result models that are particularly insensitive to detector effects.
Adding a further Lund-plane angular cut $\ln 1/\Delta < 4$ to remove
unmitigated effects due to electromagnetic calorimeter granularity did
not provide any noticeable improvement, as is shown in dashed lines in
the figure for the Lund+LSTM model.

The limitations in achieving higher resilience values for any of the
considered models are due to the consistently enhanced performance of
taggers at hadron-level.
We show the ROC curves for each model in figure~\ref{fig:roc-detec}
(left), with the dotted lines showing the hadron-level model applied
on detector-level data.
Here one can note the performance of the Lund+LSTM model on the
detector-level sample, achieving performance quite close to the
LundNet and ParticleNet models.
The LundNet-5 model in particular, is performing slightly worse than
LundNet-3 when applied on the detector-level sample, despite having a
substantial edge over it on the hadron-level data that both were
trained on.
The lower panel gives the ratio between both curves, with the
background rejection ratio of the Lund+LSTM tagger with an angular cut
$\ln 1/\Delta<4$ shown in dashed lines.
In the right-hand side of figure~\ref{fig:roc-detec}, one can see the
ROC curve of the \mbox{LundNet-3} model trained for increasing $k_t$ cuts,
showing somewhat improved robustness at larger cut values.

\subsection{Complexity of models}
\label{sec:complexity}
An important quality for a deep learning-based jet tagger is the
simplicity of the model, and the speed of its training and inference on new samples.
To quantify these considerations we measure three different metrics for the models:%
\footnote{We do not include a comparison with RecNN as this model was trained on a CPU.}
\begin{itemize}
\item the number of trainable parameters of the model,
\item the training time of each model per data sample and per epoch,
  which provides a measure of the time needed to train a full tagger
  on a given data set,
\item and finally the inference time per sample of the trained model
  on new data points, which provides a measure of how efficiently an existing
  model can be deployed to label a given sample of jets.
\end{itemize}

\begin{table}
  \centering
  \begin{tabular}{cccccc}
    \toprule
     && Number of & & Training time & Inference time\\
     && parameters & & [ms/sample/epoch] & [ms/sample]\\
    \midrule
    LundNet && 395k  &&  0.472  & 0.117 \\[4pt]
    ParticleNet && 369k && 3.488  & 1.036 \\[4pt]
    Lund+LSTM && 67k && 0.424 & 0.131 \\[2pt]
    \bottomrule
  \end{tabular}
  \caption{Summary for each model of the number of parameters,
    training time per sample and epoch, and inference time per
    sample.
    The time is measured in milliseconds as obtained when running the
    models on an Nvidia GTX 1080 Ti card.
  }
  \label{tab:model-comparison}
\end{table}

We evaluate the training time on the 2 TeV $W$ and QCD training sample
used previously in section~\ref{sec:wtag}, and the inference time on
the corresponding test data of 100k jets.
The results are shown in table~\ref{tab:model-comparison} for the 
graph-based models and the LSTM tagger.
As \mbox{LundNet-3} and \mbox{LundNet-5} only differs in the dimension of the input
features, the number of parameters and the computational cost are
essentially the same, therefore we do not distinguish between them in
this section and provide numbers derived from the \mbox{LundNet-3} tagger.
The Lund+LSTM model has a much simpler architecture, resulting in only
67k trainable parameters, significantly less than any of the
graph-based models.
It is however not substantially faster than these larger models, and
even underperforms the LundNet models in inference time.
The relatively long training time is partly due to the smaller
learning rate used when training the LSTM network, and the smaller
number of epochs needed for the Lund+LSTM model to converge.
Due to its increased number of EdgeConv blocks, the LundNet model
has 26k more parameters than ParticleNet.
However, the direct use of the Lund tree as the graph structure
removes the need for a costly nearest-neighbour search and also 
significantly reduces the number of edges for each node, therefore
increasing both the training and inference speed by almost an order of
magnitude.
This is compounded by the fact that due to their higher-level
kinematic inputs, the LundNet models take significantly less epochs to
converge to a good solution.%
\footnote{We note that in this benchmark the time needed to
  pre-process jets from list of particles to input data to each model
  is not included.
  Due to its reliance on recursion, our python implementation takes
  about 4.3 ms per jet to recluster a jet and transform the clustering
  tree into a graph of Lund nodes.
  This is however completely dependent on the data format used when
  saving \texttt{Pythia} events and can be therefore significantly
  reduced through a more efficient processing pipeline
  implementation.}

\begin{figure}
  \centering
  \includegraphics[width=0.8\textwidth,page=1]{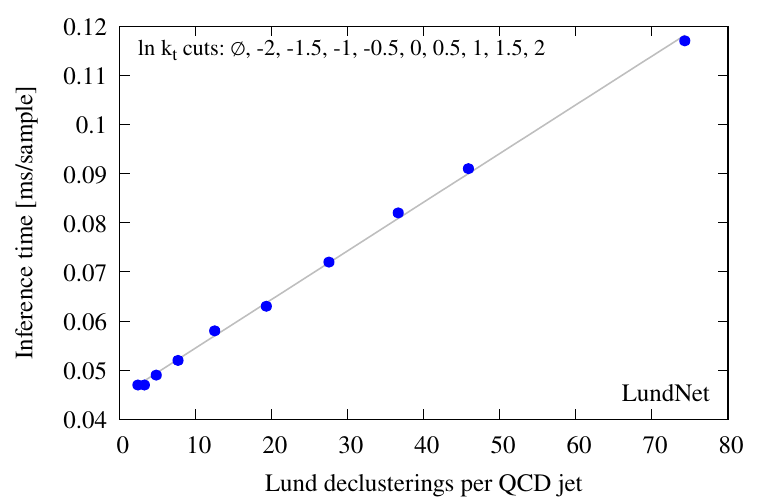}
  \caption{Inference time per jet of the LundNet model as a function
    of the mean number of Lund declusterings per 2 TeV QCD jet.
    Each circle corresponds to a separate LundNet model trained for a
    different $k_t$ cut, as indicated in the figure text.}
  \label{fig:inference-time}
\end{figure}

An interesting side-effect of the $k_t$ cut applied in the Lund plane
to improve the robustness of the model as described in previous
sections is that it also reduces the number of nodes present in the
graph.
As such, both the training and the inference time of the model are
expected to be reduced as the transverse momentum cut is increased and
more nodes are removed from the input graph.
To demonstrate this we show in figure~\ref{fig:inference-time} the
inference time per sample as a function of the average number of Lund
declusterings per QCD jet, obtained through models trained with
different Lund plane $k_t$ cuts, each of which is shown as blue
circle.
As expected, the inference time scales linearly with the number of
nodes in the graph, such that computing time increases quadratically
as the $\ln k_t$ cut is reduced.

\section{Conclusions}
\label{sec:concl}

In this article, we have introduced LundNet, a novel algorithm used to
detect signals at the LHC.
We showed that this method provides substantial improvements over
existing methods on the identification of key benchmark processes, as
well as in training speed and robustness to non-perturbative effects.

The LundNet model combines the power of graph convolutional networks
with an efficient representation of the radiation patterns within a
jet to optimally extract information from its substructure.
Jets are represented through a Lund tree constructed from the CA
clustering tree of each jet.
Each node of the Lund tree contains a tuple of kinematic information
for the corresponding pairwise splitting, used as input to the graph
network.
By using the clustering tree structure to aggregate information in the
graph convolution, the weights of the LundNet model can be trained ten
times faster than previous graph-based methods such as ParticleNet,
with a similar gain on the inference time when applying the trained
model to identify new jets.

We introduced two taggers, \mbox{LundNet-3} and \mbox{LundNet-5}
which rely on a three- and five-dimensional emission feature space
respectively.
The \mbox{LundNet-5} tagger outperforms current state-of-the-art
methods on several pivotal jet tagging benchmarks, most notably for
the identification of jets originating from top decays, while the
lower-dimensional \mbox{LundNet-3} tagger matches the performance of
current jet taggers despite its reduced kinematic input size.

We provided a concrete study of its resilience to model-dependent
non-perturbative and detector effects.
Through the use of an appropriate transverse momentum cut in the Lund
plane, we showed how one can establish an algorithm that retains high
performance while maintaining a handle on robustness.
Due to its limited kinematic input, the \mbox{LundNet-3} tagger is
best positioned to provide jet identification that is relatively
insensitive to non-perturbative effects and detector smearing, while
substantially outperforming previous methods based on the primary Lund
plane.

These results offer a concrete avenue to implementing effective
machine-learning based taggers that can be robust to model-dependent
effects present in the training data, a key feature for real-life
applications of artificial intelligence at the LHC.
In this context, the work presented in this article provides a key
step towards a new generation of efficient, robust and tractable jet
substructure tools for LHC physics.

\section*{Acknowledgments}
We are grateful to Gregory Soyez and Gavin Salam for useful
discussions and comments on the manuscript.
F.D.\ is supported by the Science and Technology Facilities Council
(STFC) under grant ST/P000770/1.
The initial stage of H.Q.'s work was performed at UC Santa Barbara 
and supported by the U.S. Department of Energy
under Grant No. DE-SC0011702.

\bibliographystyle{JHEP}
\bibliography{lund}

\providecommand{\href}[2]{#2}\begingroup\raggedright\begin{thebibliography}{10}

\bibitem{Salam:2009jx}
G.~P. Salam, {\it {Towards Jetography}},  {\em Eur. Phys. J. C} {\bf 67} (2010)
  637--686, [\href{http://arxiv.org/abs/0906.1833}{{\tt arXiv:0906.1833}}].

\bibitem{Abdesselam:2010pt}
A.~Abdesselam et~al., {\it {Boosted Objects: A Probe of Beyond the Standard
  Model Physics}},  {\em Eur. Phys. J. C} {\bf 71} (2011) 1661,
  [\href{http://arxiv.org/abs/1012.5412}{{\tt arXiv:1012.5412}}].

\bibitem{Altheimer:2012mn}
A.~Altheimer et~al., {\it {Jet Substructure at the Tevatron and LHC: New
  results, new tools, new benchmarks}},  {\em J. Phys. G} {\bf 39} (2012)
  063001, [\href{http://arxiv.org/abs/1201.0008}{{\tt arXiv:1201.0008}}].

\bibitem{Altheimer:2013yza}
A.~Altheimer et~al., {\it {Boosted Objects and Jet Substructure at the LHC.
  Report of BOOST2012, held at IFIC Valencia, 23rd-27th of July 2012}},  {\em
  Eur. Phys. J. C} {\bf 74} (2014), no.~3 2792,
  [\href{http://arxiv.org/abs/1311.2708}{{\tt arXiv:1311.2708}}].

\bibitem{Adams:2015hiv}
D.~Adams et~al., {\it {Towards an Understanding of the Correlations in Jet
  Substructure}},  {\em Eur. Phys. J. C} {\bf 75} (2015), no.~9 409,
  [\href{http://arxiv.org/abs/1504.00679}{{\tt arXiv:1504.00679}}].

\bibitem{Marzani:2019hun}
S.~Marzani, G.~Soyez, and M.~Spannowsky, {\em {Looking inside jets: an
  introduction to jet substructure and boosted-object phenomenology}},
  vol.~958.
\newblock Springer, 2019.

\bibitem{deOliveira:2015xxd}
L.~de~Oliveira, M.~Kagan, L.~Mackey, B.~Nachman, and A.~Schwartzman, {\it
  {Jet-images — deep learning edition}},  {\em JHEP} {\bf 07} (2016) 069,
  [\href{http://arxiv.org/abs/1511.05190}{{\tt arXiv:1511.05190}}].

\bibitem{Komiske:2016rsd}
P.~T. Komiske, E.~M. Metodiev, and M.~D. Schwartz, {\it {Deep learning in
  color: towards automated quark/gluon jet discrimination}},  {\em JHEP} {\bf
  01} (2017) 110, [\href{http://arxiv.org/abs/1612.01551}{{\tt
  arXiv:1612.01551}}].

\bibitem{Louppe:2017ipp}
G.~Louppe, K.~Cho, C.~Becot, and K.~Cranmer, {\it {QCD-Aware Recursive Neural
  Networks for Jet Physics}},  {\em JHEP} {\bf 01} (2019) 057,
  [\href{http://arxiv.org/abs/1702.00748}{{\tt arXiv:1702.00748}}].

\bibitem{Kasieczka:2017nvn}
G.~Kasieczka, T.~Plehn, M.~Russell, and T.~Schell, {\it {Deep-learning Top
  Taggers or The End of QCD?}},  {\em JHEP} {\bf 05} (2017) 006,
  [\href{http://arxiv.org/abs/1701.08784}{{\tt arXiv:1701.08784}}].

\bibitem{Butter:2017cot}
A.~Butter, G.~Kasieczka, T.~Plehn, and M.~Russell, {\it {Deep-learned Top
  Tagging with a Lorentz Layer}},  {\em SciPost Phys.} {\bf 5} (2018), no.~3
  028, [\href{http://arxiv.org/abs/1707.08966}{{\tt arXiv:1707.08966}}].

\bibitem{Larkoski:2017jix}
A.~J. Larkoski, I.~Moult, and B.~Nachman, {\it {Jet Substructure at the Large
  Hadron Collider: A Review of Recent Advances in Theory and Machine
  Learning}},  \href{http://arxiv.org/abs/1709.04464}{{\tt arXiv:1709.04464}}.

\bibitem{Cheng:2017rdo}
T.~Cheng, {\it {Recursive Neural Networks in Quark/Gluon Tagging}},  {\em
  Comput. Softw. Big Sci.} {\bf 2} (2018), no.~1 3,
  [\href{http://arxiv.org/abs/1711.02633}{{\tt arXiv:1711.02633}}].

\bibitem{Macaluso:2018tck}
S.~Macaluso and D.~Shih, {\it {Pulling Out All the Tops with Computer Vision
  and Deep Learning}},  {\em JHEP} {\bf 10} (2018) 121,
  [\href{http://arxiv.org/abs/1803.00107}{{\tt arXiv:1803.00107}}].

\bibitem{Abdughani:2018wrw}
M.~Abdughani, J.~Ren, L.~Wu, and J.~M. Yang, {\it {Probing stop pair production
  at the LHC with graph neural networks}},  {\em JHEP} {\bf 08} (2019) 055,
  [\href{http://arxiv.org/abs/1807.09088}{{\tt arXiv:1807.09088}}].

\bibitem{Moreno:2019bmu}
E.~A. Moreno, O.~Cerri, J.~M. Duarte, H.~B. Newman, T.~Q. Nguyen, A.~Periwal,
  M.~Pierini, A.~Serikova, M.~Spiropulu, and J.-R. Vlimant, {\it {JEDI-net: a
  jet identification algorithm based on interaction networks}},  {\em Eur.
  Phys. J. C} {\bf 80} (2020), no.~1 58,
  [\href{http://arxiv.org/abs/1908.05318}{{\tt arXiv:1908.05318}}].

\bibitem{Kasieczka:2019dbj}
A.~Butter et~al., {\it {The Machine Learning Landscape of Top Taggers}},  {\em
  SciPost Phys.} {\bf 7} (2019) 014,
  [\href{http://arxiv.org/abs/1902.09914}{{\tt arXiv:1902.09914}}].

\bibitem{Qu:2019gqs}
H.~Qu and L.~Gouskos, {\it {ParticleNet: Jet Tagging via Particle Clouds}},
  {\em Phys. Rev. D} {\bf 101} (2020), no.~5 056019,
  [\href{http://arxiv.org/abs/1902.08570}{{\tt arXiv:1902.08570}}].

\bibitem{Ren:2019xhp}
J.~Ren, L.~Wu, and J.~M. Yang, {\it {Unveiling CP property of top-Higgs
  coupling with graph neural networks at the LHC}},  {\em Phys. Lett. B} {\bf
  802} (2020) 135198, [\href{http://arxiv.org/abs/1901.05627}{{\tt
  arXiv:1901.05627}}].

\bibitem{Lim:2020igi}
S.~H. Lim and M.~M. Nojiri, {\it {Morphology for Jet Classification}},
  \href{http://arxiv.org/abs/2010.13469}{{\tt arXiv:2010.13469}}.

\bibitem{Datta:2017rhs}
K.~Datta and A.~Larkoski, {\it {How Much Information is in a Jet?}},  {\em
  JHEP} {\bf 06} (2017) 073, [\href{http://arxiv.org/abs/1704.08249}{{\tt
  arXiv:1704.08249}}].

\bibitem{Datta:2017lxt}
K.~Datta and A.~J. Larkoski, {\it {Novel Jet Observables from Machine
  Learning}},  {\em JHEP} {\bf 03} (2018) 086,
  [\href{http://arxiv.org/abs/1710.01305}{{\tt arXiv:1710.01305}}].

\bibitem{Lim:2018toa}
S.~H. Lim and M.~M. Nojiri, {\it {Spectral Analysis of Jet Substructure with
  Neural Networks: Boosted Higgs Case}},  {\em JHEP} {\bf 10} (2018) 181,
  [\href{http://arxiv.org/abs/1807.03312}{{\tt arXiv:1807.03312}}].

\bibitem{Komiske:2017aww}
P.~T. Komiske, E.~M. Metodiev, and J.~Thaler, {\it {Energy flow polynomials: A
  complete linear basis for jet substructure}},  {\em JHEP} {\bf 04} (2018)
  013, [\href{http://arxiv.org/abs/1712.07124}{{\tt arXiv:1712.07124}}].

\bibitem{Komiske:2018cqr}
P.~T. Komiske, E.~M. Metodiev, and J.~Thaler, {\it {Energy Flow Networks: Deep
  Sets for Particle Jets}},  {\em JHEP} {\bf 01} (2019) 121,
  [\href{http://arxiv.org/abs/1810.05165}{{\tt arXiv:1810.05165}}].

\bibitem{Chakraborty:2019imr}
A.~Chakraborty, S.~H. Lim, and M.~M. Nojiri, {\it {Interpretable deep learning
  for two-prong jet classification with jet spectra}},  {\em JHEP} {\bf 19}
  (2020) 135, [\href{http://arxiv.org/abs/1904.02092}{{\tt arXiv:1904.02092}}].

\bibitem{Kasieczka:2020nyd}
G.~Kasieczka, S.~Marzani, G.~Soyez, and G.~Stagnitto, {\it {Towards Machine
  Learning Analytics for Jet Substructure}},  {\em JHEP} {\bf 09} (2020) 195,
  [\href{http://arxiv.org/abs/2007.04319}{{\tt arXiv:2007.04319}}].

\bibitem{Agarwal:2020fpt}
G.~Agarwal, L.~Hay, I.~Iashvili, B.~Mannix, C.~McLean, M.~Morris, S.~Rappoccio,
  and U.~Schubert, {\it {Explainable AI for ML jet taggers using expert
  variables and layerwise relevance propagation}},
  \href{http://arxiv.org/abs/2011.13466}{{\tt arXiv:2011.13466}}.

\bibitem{Chakraborty:2020yfc}
A.~Chakraborty, S.~H. Lim, M.~M. Nojiri, and M.~Takeuchi, {\it {Neural
  Network-based Top Tagger with Two-Point Energy Correlations and Geometry of
  Soft Emissions}},  {\em JHEP} {\bf 20} (2020) 111,
  [\href{http://arxiv.org/abs/2003.11787}{{\tt arXiv:2003.11787}}].

\bibitem{Dolan:2020qkr}
M.~J. Dolan and A.~Ore, {\it {Equivariant Energy Flow Networks for Jet
  Tagging}},  \href{http://arxiv.org/abs/2012.00964}{{\tt arXiv:2012.00964}}.

\bibitem{lundnet_code}
F.~A. Dreyer and H.~Qu, {\it {LundNet} v1.0.0},  Jan., 2021.

\bibitem{Andersson:1988gp}
B.~Andersson, G.~Gustafson, L.~Lonnblad, and U.~Pettersson, {\it {Coherence
  Effects in Deep Inelastic Scattering}},  {\em Z. Phys.} {\bf C43} (1989) 625.

\bibitem{Dreyer:2018nbf}
F.~A. Dreyer, G.~P. Salam, and G.~Soyez, {\it {The Lund Jet Plane}},  {\em
  JHEP} {\bf 12} (2018) 064, [\href{http://arxiv.org/abs/1807.04758}{{\tt
  arXiv:1807.04758}}].

\bibitem{Aad:2020zcn}
{\bf ATLAS} Collaboration, G.~Aad et~al., {\it {Measurement of the Lund Jet
  Plane Using Charged Particles in 13 TeV Proton-Proton Collisions with the
  ATLAS Detector}},  {\em Phys. Rev. Lett.} {\bf 124} (2020), no.~22 222002,
  [\href{http://arxiv.org/abs/2004.03540}{{\tt arXiv:2004.03540}}].

\bibitem{Lifson:2020gua}
A.~Lifson, G.~P. Salam, and G.~Soyez, {\it {Calculating the primary Lund Jet
  Plane density}},  {\em JHEP} {\bf 10} (2020) 170,
  [\href{http://arxiv.org/abs/2007.06578}{{\tt arXiv:2007.06578}}].

\bibitem{Dasgupta:2020fwr}
M.~Dasgupta, F.~A. Dreyer, K.~Hamilton, P.~F. Monni, G.~P. Salam, and G.~Soyez,
  {\it {Parton showers beyond leading logarithmic accuracy}},  {\em Phys. Rev.
  Lett.} {\bf 125} (2020), no.~5 052002,
  [\href{http://arxiv.org/abs/2002.11114}{{\tt arXiv:2002.11114}}].

\bibitem{Dokshitzer:1997in}
Y.~L. Dokshitzer, G.~D. Leder, S.~Moretti, and B.~R. Webber, {\it {Better jet
  clustering algorithms}},  {\em JHEP} {\bf 08} (1997) 001,
  [\href{http://arxiv.org/abs/hep-ph/9707323}{{\tt hep-ph/9707323}}].

\bibitem{Wobisch:1998wt}
M.~Wobisch and T.~Wengler, {\it {Hadronization corrections to jet
  cross-sections in deep inelastic scattering}},  in {\em {Monte Carlo
  generators for HERA physics. Proceedings, Workshop, Hamburg, Germany,
  1998-1999}}, pp.~270--279, 1998.
\newblock \href{http://arxiv.org/abs/hep-ph/9907280}{{\tt hep-ph/9907280}}.

\bibitem{Andrews:2018jcm}
H.~A. Andrews et~al., {\it {Novel tools and observables for jet physics in
  heavy-ion collisions}},  \href{http://arxiv.org/abs/1808.03689}{{\tt
  arXiv:1808.03689}}.

\bibitem{Sjostrand:2014zea}
T.~Sjöstrand, S.~Ask, J.~R. Christiansen, R.~Corke, N.~Desai, P.~Ilten,
  S.~Mrenna, S.~Prestel, C.~O. Rasmussen, and P.~Z. Skands, {\it {An
  Introduction to PYTHIA 8.2}},  {\em Comput. Phys. Commun.} {\bf 191} (2015)
  159--177, [\href{http://arxiv.org/abs/1410.3012}{{\tt arXiv:1410.3012}}].

\bibitem{DGCNN}
Y.~Wang, Y.~Sun, Z.~Liu, S.~E. Sarma, M.~M. Bronstein, and J.~M. Solomon, {\it
  Dynamic graph cnn for learning on point clouds},  {\em ACM Trans. Graph.}
  {\bf 38} (Oct., 2019) 146.

\bibitem{DBLP:journals/corr/IoffeS15}
S.~Ioffe and C.~Szegedy, {\it Batch normalization: Accelerating deep network
  training by reducing internal covariate shift},  in {\em Proceedings of the
  32nd International Conference on Machine Learning}, vol.~37, (Lille, France),
  pp.~448--456, PMLR, 07--09 Jul, 2015.

\bibitem{glorot2011deep}
X.~Glorot, A.~Bordes, and Y.~Bengio, {\it Deep sparse rectifier neural
  networks},  in {\em Proceedings of the Fourteenth International Conference on
  Artificial Intelligence and Statistics}, vol.~15, (Fort Lauderdale, FL, USA),
  pp.~315--323, PMLR, 11--13 Apr, 2011.

\bibitem{he2016deep}
K.~{He}, X.~{Zhang}, S.~{Ren}, and J.~{Sun}, {\it Deep residual learning for
  image recognition},  in {\em 2016 IEEE Conference on Computer Vision and
  Pattern Recognition (CVPR)}, (Las Vegas, NV, USA), pp.~770--778, IEEE, 2016.

\bibitem{wang2019dgl}
M.~Wang, D.~Zheng, Z.~Ye, Q.~Gan, M.~Li, X.~Song, J.~Zhou, C.~Ma, L.~Yu,
  Y.~Gai, T.~Xiao, T.~He, G.~Karypis, J.~Li, and Z.~Zhang, {\it Deep graph
  library: A graph-centric, highly-performant package for graph neural
  networks},  {\em arXiv preprint arXiv:1909.01315} (2019).

\bibitem{NEURIPS2019_9015}
A.~Paszke, S.~Gross, F.~Massa, A.~Lerer, J.~Bradbury, G.~Chanan, T.~Killeen,
  Z.~Lin, N.~Gimelshein, L.~Antiga, A.~Desmaison, A.~Kopf, E.~Yang, Z.~DeVito,
  M.~Raison, A.~Tejani, S.~Chilamkurthy, B.~Steiner, L.~Fang, J.~Bai, and
  S.~Chintala, {\it Pytorch: An imperative style, high-performance deep
  learning library},  in {\em Advances in Neural Information Processing Systems
  32} (H.~Wallach, H.~Larochelle, A.~Beygelzimer, F.~d\textquotesingle
  Alch\'{e}-Buc, E.~Fox, and R.~Garnett, eds.), pp.~8024--8035.
\newblock Curran Associates, Inc., 2019.

\bibitem{DBLP:journals/corr/KingmaB14}
D.~P. Kingma and J.~Ba, {\it Adam: {A} method for stochastic optimization},
  {\em CoRR} {\bf abs/1412.6980} (2014)
  [\href{http://arxiv.org/abs/1412.6980}{{\tt arXiv:1412.6980}}].

\bibitem{Carrazza:2019efs}
S.~Carrazza and F.~A. Dreyer, {\it {Jet grooming through reinforcement
  learning}},  {\em Phys. Rev. D} {\bf 100} (2019), no.~1 014014,
  [\href{http://arxiv.org/abs/1903.09644}{{\tt arXiv:1903.09644}}].

\bibitem{groomRL_data}
S.~Carrazza and F.~A. Dreyer, {\it {JetsGame}/data v1.0.0},  Mar., 2019.
\newblock This repository is git-lfs.

\bibitem{Cacciari:2008gp}
M.~Cacciari, G.~P. Salam, and G.~Soyez, {\it {The Anti-k(t) jet clustering
  algorithm}},  {\em JHEP} {\bf 04} (2008) 063,
  [\href{http://arxiv.org/abs/0802.1189}{{\tt arXiv:0802.1189}}].

\bibitem{Cacciari:2011ma}
M.~Cacciari, G.~P. Salam, and G.~Soyez, {\it {FastJet User Manual}},  {\em Eur.
  Phys. J.} {\bf C72} (2012) 1896, [\href{http://arxiv.org/abs/1111.6097}{{\tt
  arXiv:1111.6097}}].

\bibitem{Gallicchio:2011xq}
J.~Gallicchio and M.~D. Schwartz, {\it {Quark and Gluon Tagging at the LHC}},
  {\em Phys. Rev. Lett.} {\bf 107} (2011) 172001,
  [\href{http://arxiv.org/abs/1106.3076}{{\tt arXiv:1106.3076}}].

\bibitem{Larkoski:2013eya}
A.~J. Larkoski, G.~P. Salam, and J.~Thaler, {\it {Energy Correlation Functions
  for Jet Substructure}},  {\em JHEP} {\bf 06} (2013) 108,
  [\href{http://arxiv.org/abs/1305.0007}{{\tt arXiv:1305.0007}}].

\bibitem{Larkoski:2014pca}
A.~J. Larkoski, J.~Thaler, and W.~J. Waalewijn, {\it {Gaining (Mutual)
  Information about Quark/Gluon Discrimination}},  {\em JHEP} {\bf 11} (2014)
  129, [\href{http://arxiv.org/abs/1408.3122}{{\tt arXiv:1408.3122}}].

\bibitem{Bhattacherjee:2015psa}
B.~Bhattacherjee, S.~Mukhopadhyay, M.~M. Nojiri, Y.~Sakaki, and B.~R. Webber,
  {\it {Associated jet and subjet rates in light-quark and gluon jet
  discrimination}},  {\em JHEP} {\bf 04} (2015) 131,
  [\href{http://arxiv.org/abs/1501.04794}{{\tt arXiv:1501.04794}}].

\bibitem{Frye:2017yrw}
C.~Frye, A.~J. Larkoski, J.~Thaler, and K.~Zhou, {\it {Casimir Meets Poisson:
  Improved Quark/Gluon Discrimination with Counting Observables}},  {\em JHEP}
  {\bf 09} (2017) 083, [\href{http://arxiv.org/abs/1704.06266}{{\tt
  arXiv:1704.06266}}].

\bibitem{Metodiev:2018ftz}
E.~M. Metodiev and J.~Thaler, {\it {Jet Topics: Disentangling Quarks and Gluons
  at Colliders}},  {\em Phys. Rev. Lett.} {\bf 120} (2018), no.~24 241602,
  [\href{http://arxiv.org/abs/1802.00008}{{\tt arXiv:1802.00008}}].

\bibitem{Larkoski:2019nwj}
A.~J. Larkoski and E.~M. Metodiev, {\it {A Theory of Quark vs. Gluon
  Discrimination}},  {\em JHEP} {\bf 10} (2019) 014,
  [\href{http://arxiv.org/abs/1906.01639}{{\tt arXiv:1906.01639}}].

\bibitem{Bendavid:2018nar}
J.~R. Andersen et~al., {\it {Les Houches 2017: Physics at TeV Colliders
  Standard Model Working Group Report}},  2018.
\newblock \href{http://arxiv.org/abs/1803.07977}{{\tt arXiv:1803.07977}}.

\bibitem{deFavereau:2013fsa}
{\bf DELPHES 3} Collaboration, J.~de~Favereau, C.~Delaere, P.~Demin,
  A.~Giammanco, V.~Lemaître, A.~Mertens, and M.~Selvaggi, {\it {DELPHES 3, A
  modular framework for fast simulation of a generic collider experiment}},
  {\em JHEP} {\bf 02} (2014) 057, [\href{http://arxiv.org/abs/1307.6346}{{\tt
  arXiv:1307.6346}}].

\bibitem{Son:2012mb}
M.~Son, C.~Spethmann, and B.~Tweedie, {\it {Diboson-Jets and the Search for
  Resonant Zh Production}},  {\em JHEP} {\bf 08} (2012) 160,
  [\href{http://arxiv.org/abs/1204.0525}{{\tt arXiv:1204.0525}}].

\end{thebibliography}\endgroup

\end{document}